\documentclass[journal]{IEEEtran}

\usepackage{amsmath,amsfonts, amssymb}
\usepackage{algorithmic}
\usepackage{algorithm}
\usepackage{array}
\usepackage[caption=false,font=normalsize,labelfont=sf,textfont=sf]{subfig}
\usepackage{textcomp}
\usepackage{stfloats}
\usepackage{url}
\usepackage{verbatim}
\usepackage{graphicx}
\graphicspath {{./img}}
\usepackage{cite}
\usepackage{graphicx}
\usepackage{cuted}
\usepackage{capt-of}
\usepackage{booktabs}
\usepackage{multirow}
\usepackage{siunitx}
\usepackage{xspace}
\usepackage{enumitem}
\usepackage{hyperref}
\usepackage{microtype}
\usepackage{csquotes}
\usepackage[dvipsnames]{xcolor}

\newcommand{\brix}{\ensuremath{^\circ}Bx\xspace}

\newcommand{\YEL}{\textcolor{YellowOrange}{YEL}}
\newcommand{\GRN}{\textcolor{OliveGreen}{GRN}}
\newcommand{\BLU}{\textcolor{MidnightBlue}{BLU}}
\newcommand{\RED}{\textcolor{BrickRed}{RED}}
\newcommand{\PUR}{\textcolor{Plum}{PUR}}

\newif\ifinpaper
\inpapertrue
\newcommand{\inpaper}[2]{\ifinpaper #1\else #2\fi}

\setlist[itemize]{leftmargin=*, topsep=2pt, itemsep=2pt}
\setlist[enumerate]{leftmargin=*, topsep=2pt, itemsep=2pt}

\begin{document}

\title{Sugar Shack 4.0: Implementation of a Cyber-Physical System for Logistic and Sanitary Automation in a Maple Syrup Boiling Center}

\author{Thomas Bernard, \IEEEmembership{Graduate Student Member, IEEE}, François Grondin, \IEEEmembership{Member, IEEE}, and Jean-Michel Lavoie
\thanks{This work was supported in part by the Natural Sciences and Engineering Research Council of Canada (NSERC) and the Consortium de recherche et innovations en bioprocédés industriels au Québec (CRIBIQ). \textit{Corresponding author: Thomas Bernard.}}
\thanks{Thomas Bernard and François Grondin are with the Department of Electrical and Computer Engineering, Université de Sherbrooke, Sherbrooke, QC J1K 2R1, Canada. (e-mail: thomas.bernard@usherbrooke.ca, francois.grondin2@usherbrooke.ca).}
\thanks{Jean-Michel Lavoie is with the Department of Chemical and Biotechnological Engineering, Université de Sherbrooke, Sherbrooke, QC J1K 2R1, Canada. (e-mail: jean-michel.lavoie2@usherbrooke.ca).}}

\maketitle

\begin{abstract}
  This paper presents the design and deployment of a process-aware cyber-physical system that automates plant-level logistics, traceability, and sanitation in a centralized maple-syrup boiling center. The system replaces ad-hoc, manual operations with event-driven orchestration on a local server, employing reusable device abstractions and a centralized interlock with priority-based arbitration for shared piping. It implements deterministic routines for delivery, reverse osmosis integration, evaporator feed, and permeate management. The system is sensor rich: inline measurements of flow, temperature, and sugar concentration (degrees Brix) drive routing decisions and trigger systematic post-transfer rinses (cleaning-in-place), ensuring consistent hygiene and complete, immediate traceability up to the evaporator inlet. During the 2025 production season, the system queued 431 operations without incident; executed 908 “topstock” and 296 “downstock” balancing cycles; increased usable permeate reserves from 22,712 to approximately 41,640 L through dynamic storage assignment; eliminated mid-season contamination incidents previously observed under manual practice; and reduced administrative effort for billing and reporting from more than 30 hours to roughly 1 hour through automatic documentation. These results demonstrate a practical path to modular, plant-scale automation beyond traditional architectures, and lay the groundwork for packaging reusable elements for similar plants or adjacent industries. This work is part of a larger project involving the first scientifically-documented integration of Industry 4.0 technologies in a maple syrup boiling center.
\end{abstract}


\section{Introduction}
  \inpaper{\IEEEPARstart{W}{ith}}{With}
  seasonal activity concentrating labor and capital into a few intense weeks each spring, maple syrup production holds an important cultural and economic role in northeastern North America, particularly in Québec which accounted for 72.1\% of the worldwide production in 2024~\cite{producteursetproductricesacericolesduquebecStatistiquesAcericoles20242024}. 
  Producers face structural pressures: persistent labor shortages and succession challenges~\cite{emploietdeveloppementsocialcanadaProfilSectorielAgriculture2024}, physically demanding seasonal work (often in remote environments), and rising equipment costs across the chain (collection systems, reverse osmosis (RO), evaporators, filter presses, etc.). These stresses incentivize operational models that can scale while reducing human error without increasing staffing requirements.

\begin{figure}[h]
    \centering
    \includegraphics[width=\inpaper{\linewidth}{0.8\linewidth}]{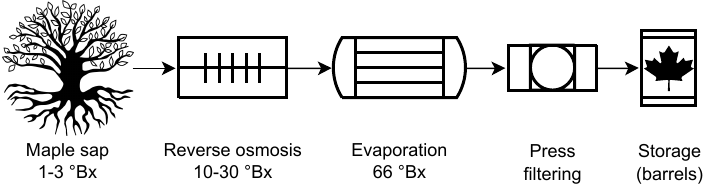}
    \caption{Modern maple processing steps.}
    \label{fig:process-overview}
  \end{figure}

  Fig.~\ref{fig:process-overview} shows the modern path from raw sap to processed syrup: sap is collected by vacuum systems in a pumping station, concentrated by RO, boiled into syrup by an evaporator, filtered, and stored at pasteurizing temperature in barrels. In practice, each step entails multiple operations, shared piping, and strict hygiene routines that must be executed consistently to avoid contamination while preserving quality. 
  Similarly to how the dairy industry sets pricing of raw material based on fat contents, the maple syrup industry's pricing rules revolve around sugar contents, measured in degrees Brix (\brix, \%~dissolved sugars; often measured by refractometry), with a set price for syrup at 66 \brix~\cite{producteursetproductricesacericolesduquebecStatistiquesAcericoles20242024}. This means a combined \brix and volume measurement can be used to attribute a price to raw material at any concentration.

  \inpaper{\IEEEpubidadjcol}{}

  A rapidly growing practice to address the maple syrup industry's challenges is to centralize processing in "boiling centers". These facilities decouple the sugarbush from transformation operations by purchasing bulk raw material from the producers, who can then focus their efforts on sap collection operations (and optionally a first RO pass to save on transportation costs)~\cite{legisquebecReglementContingentementProducteurs2025}. This model allows economies of scale but introduces new constraints: (i) high infrastructure complexity due to inbound variability (concentration, temperature, quality, timing), carrying a higher risk of operator error and shared-resource conflicts; (ii) rigorous traceability requirements because materials from different producers may mix and need to be individually tracked beforehand; and (iii) high importance of sanitation practices due to shared infrastructure elevating the risk (and consequences) of contamination. 

  Scientific literature on the maple industry skews toward the \emph{forest} (tree physiology, climate effects, sap yield, vacuum collection)~\cite{lagaceEffectNewHigh2019, caughronChangingClimateMaple2021, wangMapleSyrupIndustry2025}, and commercial innovation into Internet of Things (IoT) shares this area of focus (vacuum monitoring, pumping station remote control) since such technologies are applicable to all producers, regardless of their size. While RO systems and evaporators have advanced commercially with PLC-based automation, there is a research gap around \emph{plant-level} automation coordinating multiple units, arbitrating shared piping, and automating logistics/cleaning at scale.

  To bridge this gap, related work from adjacent liquid-food domains is examined. Cleaning-in-place (CIP) is increasingly treated as a controls/data problem where consistent execution directly reduces contamination risk. In fluid milk, interventions that tightened process hygiene reduced post-pasteurization contamination (PPC) and extended shelf life across multiple commercial plants~\cite{reichlerInterventionsDesignedControl2020}; subsequent multi-plant work links PPC patterns to spoilage trajectories and quality losses~\cite{lottGramnegativePostpasteurizationContamination2024}. 
  Reviews from dairy processing synthesize that effective CIP prevents microbial attachment and biofilm establishment in tanks, heat exchangers, and piping; directly addressing the mechanism behind episodic contamination~\cite{lapointeBiofilmFormationDairy2025}. 
  The same mechanism is observed in maple processing: variations in sap microbiota across processing periods and sites are associated with the \emph{ropy} defect in syrup~\cite{filteauMapleSapPredominant2012}. Bacterial proliferation induce biofilms which adherence is increased by residence time and higher temperatures~\cite{perryComprehensiveReviewMaple2022}. 
  Although most studies characterize biofilms in sap collection tubing, the same dynamics plausibly extend to shared plant piping and storage infrastructure, where stagnant, sugar-rich films can seed downstream contamination, especially when exposed to warm ambient temperature.
  Taken together, the literature supports the use of systematic, content-aware rinses after every transfer (rather than ad-hoc, operator-dependent cleaning) as a primary lever to prevent contamination and the associated product losses; analogous arguments appear in brewing quality control, where earlier contamination mitigation is explicitly tied to reduced financial loss~\cite{oldhamMethodsDetectionIdentification2023}.
  
  On another front, inline sensing is now commonplace in liquid foods, with both wide offerings from industrial instrumentation manufacturers and research interest. In maple processing specifically, a multi-modal inline platform demonstrated continuous \brix{} tracking (impedance-\brix{} fit \(R^2{=}0.895\) with temperature offset) and optical grade sensing (linear fit \(R^2{=}0.963\)) in production lines, while also documenting electrochemical approaches were unstable at process temperatures~\cite{landariMultiModalSensingPlatform2021}. Inline \brix{}/composition sensing is also broadly reported across beverages~\cite{jaywantSensorsInstrumentsBrix2022}.
  These sensing applications serve as the basis for existing unit-level automation (e.g., PLC-based fermentation control~\cite{xiaodongDesignImplementationControl2015}) and low-cost retrofits in process industries (e.g., ZigBee monitoring of sugar melters~\cite{saravananZigbeeBasedMonitoring2014}) illustrating mature building blocks for instrumentation even when plant-level logistics are manual.

  Liquid-food production planning/scheduling provides precedent for multi-asset coordination. A recent review synthesizes dairy/brewery formulations and CIP/changeover considerations in MILP-based plans, showing that optimized plans can reduce modeled production costs by up to 29.5\% relative to baseline~\cite{samouilidouFoodProductionScheduling2023}. However, these tools are offline and assumption-driven, and plans are deployed manually rather than with a direct tie to control systems, which limits real-time response and maintains the need for manual validation and intervention.

  Addressing this limitation, several liquid-food studies frame plant automation explicitly as a cyber-physical system (CPS).
  In beverage bottling, an agent-based CPS demonstration integrated with a tight coupling to line equipment to enable flexible orchestration, yet keep real-time safety loops local to machine PLCs and stop at single-line-level coordination~\cite{marschallDesignInstallationAgentControlled2021}. 
  Complementary CPS work combines inline sensors with cloud analytics to compute real-time water-footprint metrics~\cite{cuiCyberPhysicalSystemCPS2021} and secures food traceability via fog nodes~\cite{awanFogComputingBasedCyberPhysical2022}, while remaining largely decoupled from plant automation.
  Taken together, current CPS work in adjacent liquid-food industries either (i) couples tightly to specific unit operations without generalizing to facility-wide resource arbitration, or (ii) analyzes/records process states without closing the loop on shared-infrastructure operations. 
  This leaves an opportunity for a CPS implementation that directly encapsulates automation, where content awareness, deterministic transitions, and hygiene assurance must be enforced at the plant level.

  As part of a larger project aiming to optimize the productivity and efficiency of a maple syrup boiling center using Industry 4.0 technologies, this work builds upon another paper that generalized the lower-level IIoT control architecture employed for the project~\cite{bernardSugarShack402025}.
  The present paper contributes the practical application of said architecture as a process-aware CPS in the boiling center, piloting plant-level logistics, traceability and hygiene. 
  The remainder of the paper is structured as follows: Section \ref{sec:arch} details the system architecture and its implementation as concurrent subsystems, and Section \ref{sec:outcomes} presents and discusses the operational outcomes of the system as part of the production season of 2025.

\section{System architecture and implementation}
\label{sec:arch}
  This section presents the deployed CPS that supervises plant-level logistics, traceability, and CIP across a maple syrup boiling center. The scope of integration at the time of writing is from raw material arrival up to the evaporator inlet. Within this scope, the CPS has 100\% process coverage; no manual actuation or data collection is performed, all interactions are performed via Human-Machine Interface (HMI).
  
  The system implements a fully event-driven control layer orchestrated in Node-RED~\cite{LowcodeProgrammingEventdriven}, that models plant behavior as \emph{operations} (short, composable actions) which are launched by \emph{routines} (subsystem state-machines). 
  Devices (valves, pumps, sensors) are abstracted over MQTT~\cite{MQTTStandardIoT}, allowing control logic to remain hardware-agnostic and reusable across subsystems. 
  A centralized interlock with priority queueing arbitrates shared resources, so simultaneous operations are safe without enumerating conflict cases at design time. 
  Full architectural details (timing, abstractions, interlock, safety considerations) are documented in~\cite{bernardSugarShack402025}; hence the work presented in this paper benefits from two premises: (i) device-level integration is transparent and (ii) no specific considerations need to be presented for ensuring the safe concurrent execution of subsystem operations.
  HMI is implemented with Node-RED dashboard, accessed on fixed-position Raspberry Pi~\cite{RaspberryPi} stations throughout the plant and on mobile devices.

  This section first presents plant-level context and system components that are available to all subsystems in Section~\ref{subsec:dynamicsilos}.
  The remainder of the section is ordered by subsystem. Each one presents, where applicable: role, relevant algorithms, HMI workflow and traceability implications:
  Section~\ref{subsec:delivery} details delivery logistics; 
  Section~\ref{subsec:ro} integrates the RO system;
  Section~\ref{subsec:evap} covers evaporator feed control; 
  Section~\ref{subsec:permeate} completes with permeate balancing to improve available reserves and asset utilization.
  
  \subsection{Plant-wide context: shared assets, content model, routing}
  \label{subsec:dynamicsilos}

    The deployed CPS is content-aware, meaning that content type, level and temperature within all storage infrastructure are tracked at all times. The system supports four content types: sap, concentrate, permeate\footnote{Permeate is demineralized water, obtained as the byproduct of RO concentration. Since local organic maple syrup regulations prohibit the use of cleaning chemicals for transit and storage infrastructure during the operating season, permeate is extensively used for CIP operations.} and exception\footnote{Exception is used to manually tag an inbound batch of concentrate as potentially problematic, isolating it from automatic mixing with other lots (details in Section~\ref{subsec:delivery}).}. Since the current deployment extends only up to the evaporator inlet, syrup is out of scope of this work and needs not be considered as a content type.  
    The boiling center operates seven storage silos (ID~1--7) connected through a manifold to five shared transfer \emph{lines} that are arbitrated by the interlock. 
    The lines have a predefined role and are color-coded as follows: \YEL~(delivery), \GRN~(evaporator feed), \BLU~(RO concentrate output), \RED~(drain / permeate recirculation) and \PUR~(CIP).
    Five silos (ID~1--5) are \emph{dynamic}: they may be assigned at runtime to either concentrate or permeate, while two silos (ID~6--7) have fixed roles. 
    Table~\ref{tab:silos} summarizes capacities (L), roles and default ties.
    
    \begin{figure*}
      \centering
      \includegraphics[width=\linewidth]{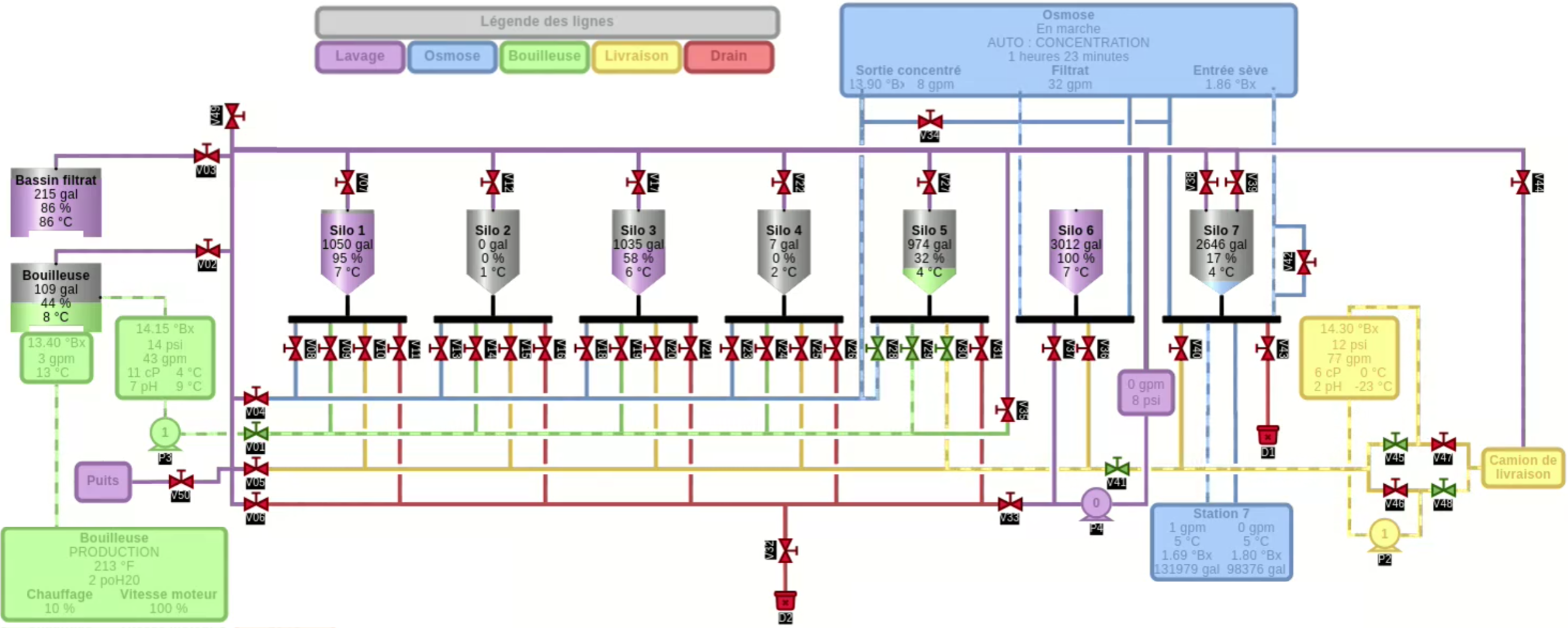}
      \captionof{figure}{\textbf{Overview HMI:} dynamic PFD of the maple syrup boiling center. Note open valves in green, active pumps marked '1', active lines with dotted lines.}
      \label{fig:pfd}
    \end{figure*}

    \begin{table}[h]
      \centering
      \caption{Silo inventory and default roles/ties.}
      \label{tab:silos}
      \renewcommand{\arraystretch}{1.15}
      \begin{tabular}{@{}rrcl@{}}
        \toprule
        \multicolumn{1}{c}{ID} &
        \multicolumn{1}{c}{Capacity} &
        \multicolumn{1}{c}{Role} &
        \multicolumn{1}{c}{Direct ties} \\
        \midrule
        1 & 4{,}164  & dynamic  & \YEL, \GRN, \BLU, \RED, \PUR~(washball) \\
        2 & 3{,}785  & dynamic  & \YEL, \GRN, \BLU, \RED, \PUR~(washball) \\
        3 & 6{,}814  & dynamic  & \YEL, \GRN, \BLU, \RED, \PUR~(washball) \\
        4 & 11{,}356 & dynamic  & \YEL, \GRN, \BLU, \RED, \PUR~(washball) \\
        5 & 11{,}356 & dynamic  & \YEL, \GRN, \BLU, \RED, \PUR~(washball) \\
        6 & 11{,}356 & permeate & \YEL, \PUR~(CIP source) / RO permeate I/O\\
        7 & 59{,}052 & sap      & \YEL, RO sap I/O, two piped-in sugarbushes \\
        \bottomrule
      \end{tabular}
    \end{table}

    While the four other lines connect to the bottom of silos as a manifold, \PUR~is plumbed to washballs at the top of each silo and as reverse-directed branches into each of the other lines. After any transfer, an automated CIP rinse with permeate can be executed along the just-used path. This both (i) recovers residual sugars and recycles them for processing and (ii) prevents stagnant sugar water, mitigating microbial growth and the risk of \emph{ropy} syrup defects. 

    The state of the complete CPS can be visualized in real time on an enhanced process flow diagram (PFD) implemented with the GoJS library and embedded in an HMI page called "Overview", as shown in Fig.~\ref{fig:pfd}. It contains:
    \begin{itemize}
      \item color-coded silo contents;
      \item sensor readings (transit measurement stations, relevant equipment PLC data, storage levels and temperatures);
      \item actuator states (valves and pumps);
      \item interlock and queue states with animations on active paths.
    \end{itemize}

    Because line access is arbitrated by the interlock, concurrent operations from different subsystems can involve the same silo. For example, Fig.~\ref{fig:pfd} shows three simultaneous operations on \textbf{Silo~5}: a concentrate delivery on \YEL, an RO concentrate output on \BLU, and an evaporator buffer feed on \GRN.

\begin{figure}[h!]
      \centering
      \includegraphics[width=\linewidth]{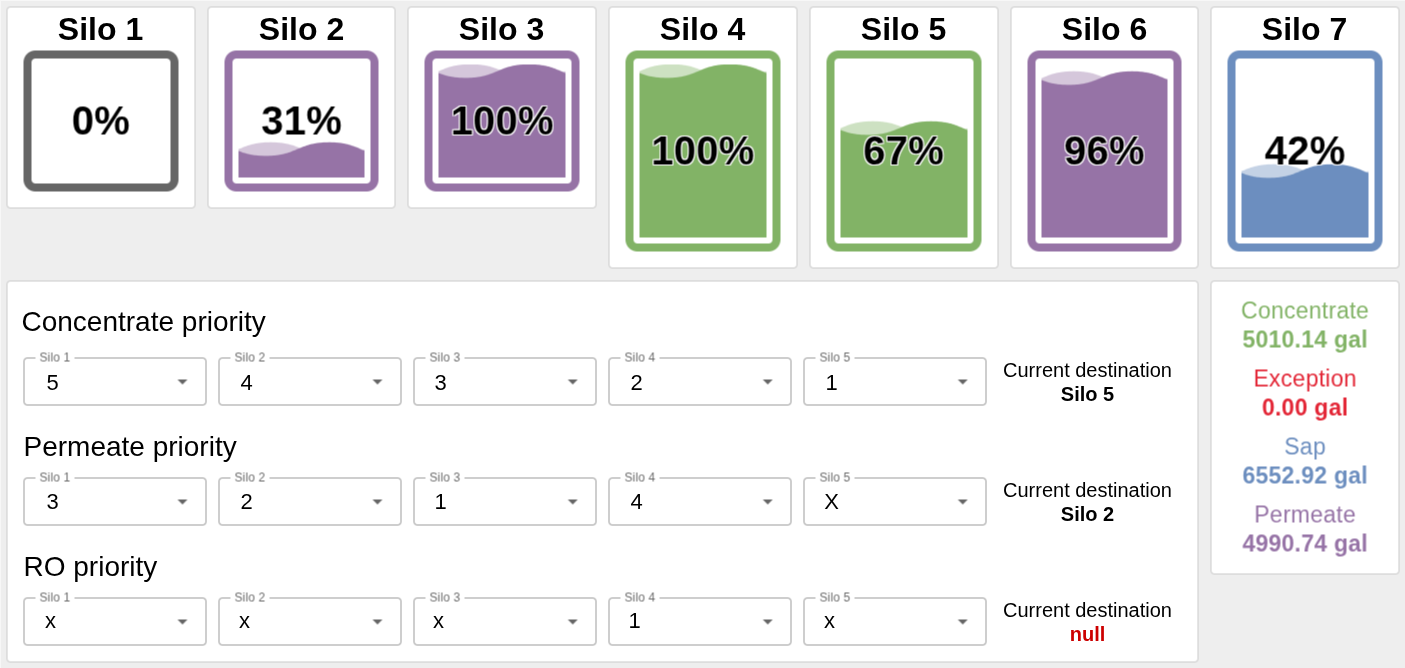}
      \caption{\textbf{Silos HMI:} live levels and content types (top); priority pickers and destinations (bottom-left); category totals (bottom-right).}
      \label{fig:hmi-silos}
    \end{figure}

    Dynamic silos raise effective utilization of existing infrastructure, while the added complexity is absorbed by policy checks from the CPS.
    Fig.~\ref{fig:hmi-silos} presents the "Silos" HMI. 
    The top row shows all silos with current levels and content types; the side panel aggregates live totals by type. For the five dynamic silos, operators define priorities via dropdowns for three routing categories: \textbf{concentrate}, \textbf{permeate}, and \textbf{RO}. For each category, a silo receives a unique rank 1-5 (1 = highest) or \texttt{X} (excluded). 
    Any change to priorities or to silo content tags recalculates the destination for the affected category. These destinations are used by the subsystems to plan routes and to surface alerts when an operation would require an unavailable target.

    Each category independently computes a destination silo using the following rules:
      
    \begin{enumerate}[label=\textbf{R\arabic*}]
      \item \textbf{Ordered scan.} Iterate dynamic silos in increasing priority (1~$\rightarrow$~5), skipping any marked \texttt{X}.
      \item \textbf{Content compatibility.} A candidate is acceptable if its current content type matches or it is empty.
      \item \textbf{Capacity check.} The candidate must have free volume (\(\text{level} < \text{threshold; e.g.,} 97\%\)).
      \item \textbf{Select or continue.} The first candidate that passes \textbf{R2}-\textbf{R3} becomes the destination for that category; if none pass, destination is set to \texttt{null}.
    \end{enumerate}

  \subsection{Delivery logistics}
  \label{subsec:delivery}

    The boiling center receives raw material from two methods: tank truck deliveries (5 origins) and directly-piped sap pumping stations (2 origins). The latter's processing is simple: the sap is pumped by external infrastructure and arrives in two distinct pipes with sufficient pressure to overcome the static pressure of silo 7 at full capacity. Each pipe has individual instruments providing continuous measurements of volume, temperature, and \brix, followed by check valves to prevent mixing or backflow. Raw data (1 Hz readings) are kept for full auditability, total respective production is updated in real time, and automatic daily reports are emailed to the producer, including cumulative data and equivalent syrup calculations. This part of the system is entirely automated and does not require any human intervention.

    In contrast, truck deliveries are more complex: arrivals are intermittent and vary in composition. The delivery station must (i) route material to a valid destination without risking overflow or cross-contamination, (ii) capture full traceability for payables and audits, and (iii) ensure the truck and transfer piping are left clean to reduce contamination risk and reclaim residual sugars. To fulfill this on the hardware side, the delivery station implements inline sensors for volume, temperature and \brix\footnote{The station also implements pH and viscosity measurements, but they are part of future work and out of scope of this paper.}, placed downstream of the delivery pump. Four valves (V45-V48 on Fig.~\ref{fig:pfd}) allow the assembly to be run in both directions (i.e., from truck to silo or from silo to truck), ensuring uninterrupted measurements during all operation steps. This station connects \YEL~to the bottom of the truck tank via a flexible tube. Another tube connects \PUR~to washballs inside the truck tank (via V44).

\begin{figure}[h]
      \centering
      \includegraphics[width=\linewidth]{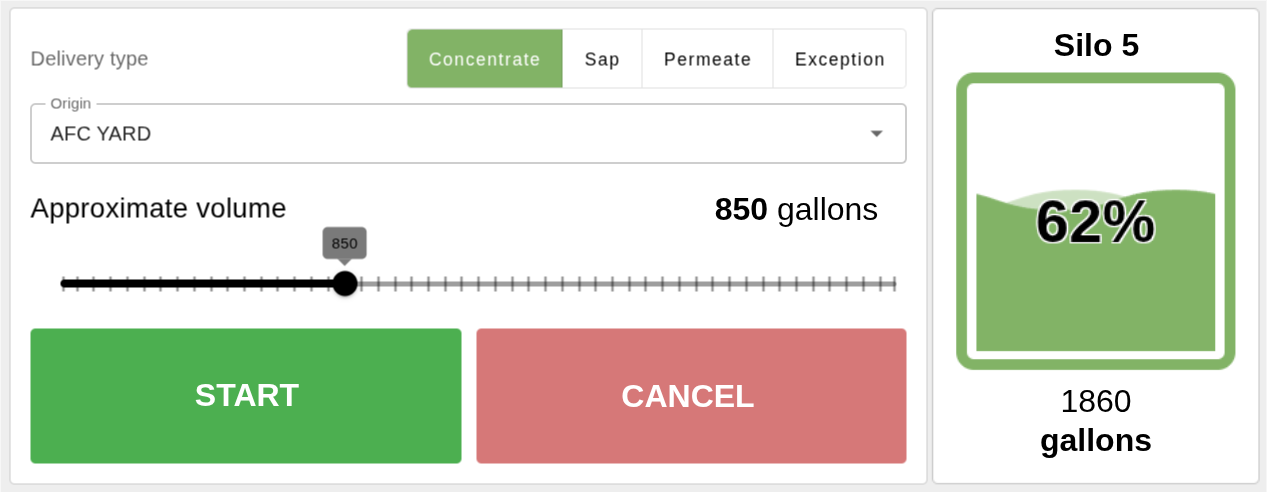} 
      \caption{\textbf{Delivery setup HMI.}}
      \label{fig:hmi-delivery-setup}
    \end{figure}

    On the software side, operators configure a delivery on the "Delivery setup" HMI (Fig.~\ref{fig:hmi-delivery-setup}): select the \textbf{delivery type} (concentrate, sap, permeate or exception), pick the \textbf{origin} from a dropdown, and set an \textbf{approximate volume} with a slider. 
    Destination is determined from the parameters set in Section~\ref{subsec:dynamicsilos}:
    \begin{itemize}
      \item Concentrate and permeate route to their computed category destinations among the five dynamic silos.
      \item Sap always routes to Silo~7.
      \item Exception prompts the operator for a destination; if confirmed, the silo's content tag is set to \emph{exception}, excluding it from automatic destinations until emptied and cleaned.
    \end{itemize}
    Every field change in the HMI recalculates the destination and shows the elected silo's label, content type and level on the right. Since the exact volume is calculated in real time during pumping, the approximate volume is only used as a guardrail to avoid starting a delivery that would not fit in the available infrastructure. Sap and exception have a single set destination, hence the system simply validates that the approximate volume is less than the available capacity in that silo. For concentrate and permeate, if that first check fails, the system checks total availability for the selected type among all dynamic silos, using the same rules as in Section~\ref{subsec:dynamicsilos} to evaluate the feasibility of a split delivery between multiple silos.
    
\begin{figure}[h]
      \centering
      \includegraphics[width=\linewidth]{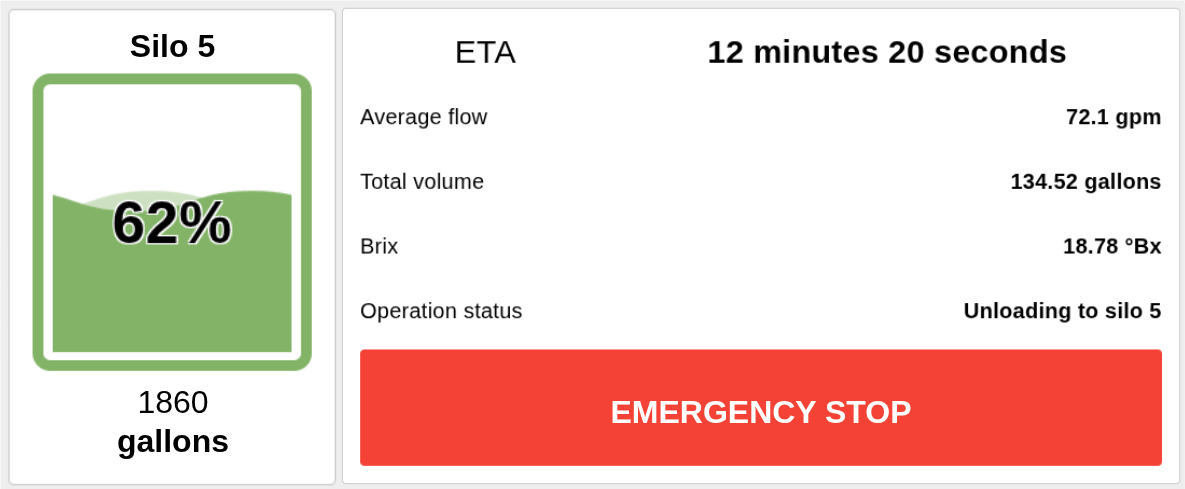}
      \caption{\textbf{Ongoing delivery HMI.}}
      \label{fig:hmi-delivery-ongoing}
    \end{figure}

    If all input fields and calculated destination are valid, a confirmation summary is displayed to reduce setup errors. When approved, the delivery sequence starts and the "Ongoing delivery" HMI (Fig.~\ref{fig:hmi-delivery-ongoing}) presents the estimated time remaining (updated from the live average flow), the state of the receiving silo, and live data: flow rate, cumulative volume, and \brix.

    Once started, the delivery routine runs the following steps:
    \begin{enumerate}
      \item \textbf{Unload to silo(s)}: open the calculated \YEL~path to the destination silo, start pumping and tag the destination silo with the relevant type; compute ETA as 
      $\text{ETA}=\max\!\big(0,\ (\hat{V}-V_{\text{now}})/\bar{Q}\big)$
      where $\hat{V}$ is the operator's approximate volume, $V_{\text{now}}$ the accumulated volume, and $\bar{Q}$ the running average flow. If the silo reaches full capacity, the pump pauses and resumes after the changeover to the next silo is completed. This changeover also sets the new destination silo's content tag.
      \item \textbf{Back-rinse to truck (concentrate/exception)}: execute a short reverse \PUR$\rightarrow$\YEL~rinse from the manifold back to the truck to reclaim sugars stranded in the line; the recovered volume/\brix is deducted from the main load.
      \item \textbf{Truck tank wash (concentrate/exception/sap)}: activate truck washballs via \PUR; after a delay, start the delivery pump to push the rinse to Silo~7 until a low \brix\ threshold is reached (e.g., \(0.1~^\circ\)Bx) indicating the tank is clean. The sugar recovered during this step is indexed to the main load \brix\ and appended to the credited total.
    \end{enumerate}

    \textbf{Traceability:}
    All signals (flow, temperature, \brix), state transitions, interlock grants, and mass-balance adjustments are logged under a unique \texttt{delivery\_id} in the time-series database. On completion, a delivery bill is generated and emailed automatically to the producer and the admin team, providing rapid feedback and an auditable record.

  \subsection{RO integration}
  \label{subsec:ro}

    The RO system is a self-contained commercial unit driven by its own PLC. Parameterization and cycle control are done either directly on the RO's internal touchscreen HMI or via a vendor-provided remote access software. Integration with our CPS is \emph{read-only} via a Modbus~TCP$\rightarrow$MQTT gateway that publishes operating data (flows, pressures, temperatures, in/out \brix, valve states, etc.) and the program's current operating mode / status. Low-level actuation, interlocks internal to the RO, and cleaning sequences are handled by the PLC; our orchestration only coordinates plant-side valves and traceability.

    The CPS maintains a single flag, \texttt{sugar\_in\_blu}, that indicates residual sugar may be present in the \BLU~line and upstream piping. Operation is as follows:

    \begin{enumerate}[leftmargin=*, itemsep=3pt, topsep=2pt]
      \item \textbf{On entry to \emph{Concentration} mode:}
        \begin{itemize}
          \item Open the manifold valve on \BLU~to the preselected destination silo from Section~\ref{subsec:dynamicsilos}.
          \footnote{Startup/priming of the RO takes several minutes, so this actuation does not impose timing constraints.}
          \item Tag destination silo as concentrate.
          \item Set \texttt{sugar\_in\_blu}~$\leftarrow$~\texttt{true}.
        \end{itemize}
      \item \textbf{On exit from any mode if \texttt{sugar\_in\_blu} is \texttt{true}:}
        \begin{itemize}
          \item Wait until the output valve from the RO is fully closed (\BLU~is decoupled).
          \item Run a short \PUR$\rightarrow$\BLU$\rightarrow$Silo~7 rinse.
          \footnote{The rinse length is fixed to push approximately three times the total volume of the affected branch; it is sufficient recover all residual sugar to Silo~7.}
          \item Clear \texttt{sugar\_in\_blu}~$\leftarrow$~\texttt{false}.
        \end{itemize}
    \end{enumerate}

    RO cleaning operations do not require plant-side actions: the RO has a dedicated permeate tie to Silo~6, providing a steady supply independent of CPS control. Nevertheless, the system continues to ingest telemetry and mode changes.

    \textbf{Traceability:}
    Each RO concentration run is tagged with a unique \texttt{ro\_batch\_id}; MQTT telemetry (flow, pressure, \brix\ in/out), mode transitions, BLU valve state, and any post-concentration flush are recorded under that ID in the time-series database for audit and mass-balance reconstruction.

  \subsection{Evaporator feed and operator assistance}
  \label{subsec:evap}
    
    The plant employs a sealed electric evaporator that uses steam compression as its primary heat source in steady state. Reaching this state is energy- and time-intensive, so the control objective is to maintain stability for as long as possible by avoiding starvation. 
    Like the RO system, the evaporator is a commercial, PLC-driven unit, which exposes \textit{read-only} production telemetry (temperature, pressures, motor \%, etc.) via Modbus~TCP\(\rightarrow\)MQTT for HMI display only.
    The evaporator inlet is gravity-fed by a hard-plumbed (no valve) \(\,\approx 946\,\)L buffer tank, which must never run dry when the evaporator is enabled. Two plant-side feeds reach the buffer:
    (i) a bottom feed from \GRN~(concentrate path), and
    (ii) a washball feed from \PUR~(permeate rinse).
    Having distinct feeds avoids dead volume effects along the \(\approx\)~40~ft run between the silo manifold and the buffer.
    The \GRN~feed has a dedicated pump with the same inline sensors as the delivery station (Section~\ref{subsec:delivery}), enabling traceability of materials deducted from silos during the boil.

\begin{figure}[h]
      \centering
      \includegraphics[width=0.92\linewidth]{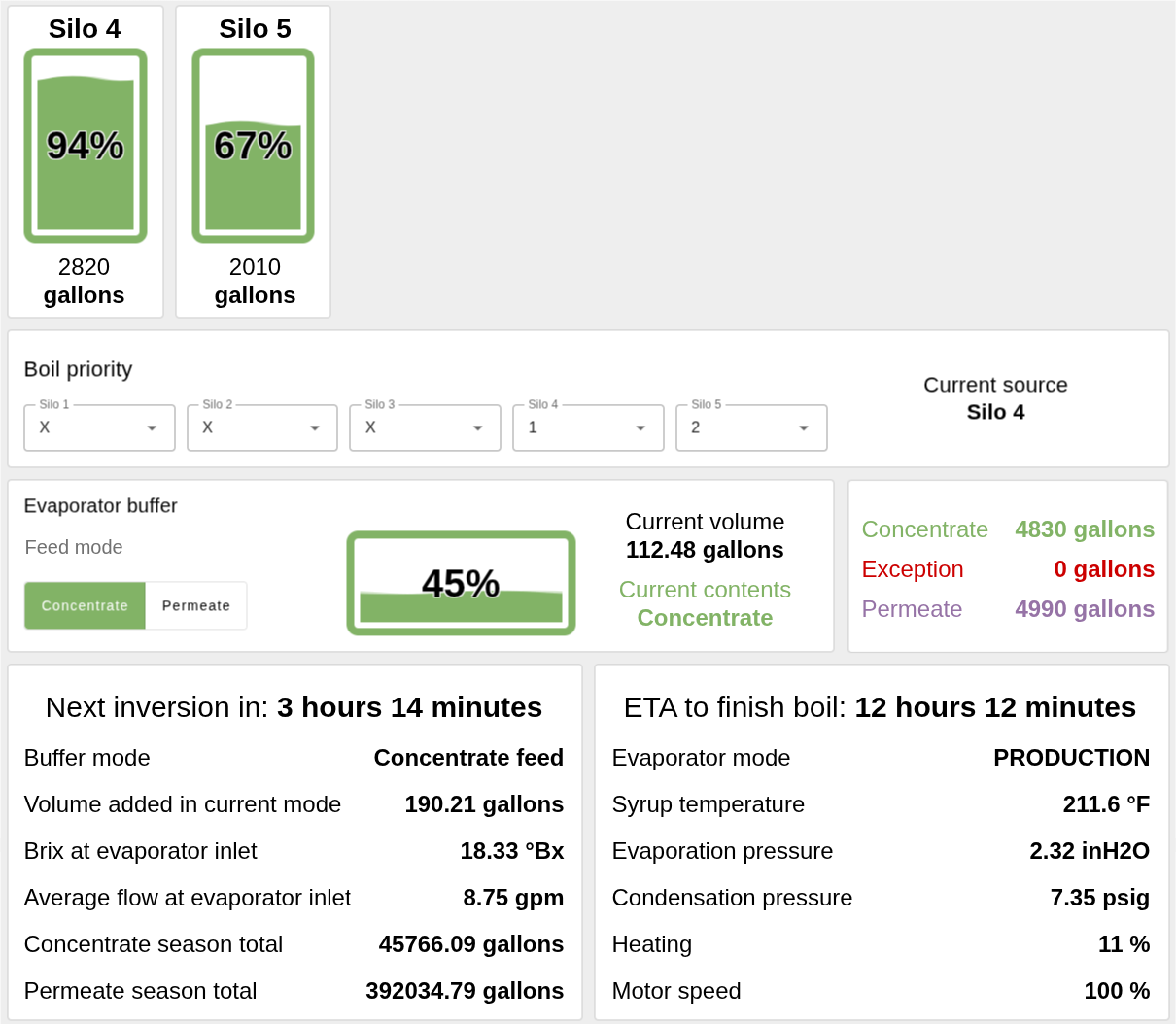}
      \caption{\textbf{Boiling HMI:} available sources and priority pickers (top); buffer contents, level, and feed mode selection (mid-left); stock totals (mid-right); and evaporator production stats (bottom).}
      \label{fig:hmi-evap-page}
    \end{figure}

    Normal buffer feed operation involves filling with either permeate or concentrate through their respective lines, depending on the operational state. The evaporator's startup sequence should always be performed with permeate, since its low thermal mass shortens the time to reach operating temperature. The "Boiling" HMI (Fig.~\ref{fig:hmi-evap-page}) allows the operator to define the boiling sequence of available silos and switch between permeate and concentrate modes. All silos tagged as concentrate or exception are listed as \textit{boiling candidates}. The operator assigns unique priorities (1 = highest) or \texttt{X} to exclude; the highest-ranked valid silo becomes the current source and the priority order defines the boil sequence.
    
    Mode changes are operator-initiated from the HMI and require a confirmation prompt;
    the CPS exposes the two following modes and corresponding transitions:
    \begin{itemize}[leftmargin=*, itemsep=2pt, topsep=2pt]
      \item \textbf{Permeate mode} (default/startup). Maintain buffer level in \([L_{\min}, L_{\max}]\) (e.g., 40-60\%) by dosing \PUR~into the washball feed whenever \(L<L_{\min}\).
      \item \textbf{Concentrate mode}. With a valid source, maintain the same band using \GRN~from the selected silo.
      \item \textbf{Transition: permeate\(\rightarrow\)concentrate}. Let the buffer fall to near-empty to avoid diluting inbound concentrate, then open \GRN~from the source; resume band control on \GRN.
      \item \textbf{Transition: concentrate\(\rightarrow\)permeate}. Let the buffer fall to near-empty, refill using \PUR; resume band control on \PUR.
    \end{itemize}
    
    Once in concentrate mode with the evaporator stable, the operator uses PLC/HMI stats and manual sample measurements to tune evaporator parameters. If \GRN~detects source exhaustion, the CPS executes a rinse with two possibilities:
    \begin{enumerate}[label=\alph*)]
      \item \textbf{Silo transition (another silo remains to boil).}
      \begin{enumerate}[label=\arabic*) ]
        \item Rinse the just-emptied silo using \PUR$\rightarrow$washball; after a delay, simultaneously push the rinse via \GRN~to the buffer until a low \brix\ threshold is reached (e.g., \(0.1~^\circ\)Bx).
        \item Close \GRN, open \RED~to drain; apply a final 30~s rinse; tag silo as empty.
        \item Select the next source from the priority list; continue concentrate mode.
      \end{enumerate}
      \item \textbf{End of boil (last concentrate silo).}
      \begin{enumerate}[label=\arabic*) ]
        \item Perform steps 1 and 2 from a).
        \item Execute a \PUR~backflush of the entire \GRN~line up to the buffer to clear any residual sugars.
        \item Automatically switch back to permeate mode.
      \end{enumerate}
    \end{enumerate}
    
    To keep internals clean, the evaporator requires a path inversion roughly every 4~hours of concentrate boiling. The HMI displays a countdown timer to the next inversion when concentrate mode is started. To avoid sugar loss, the following steps are performed by the operator:
    \begin{enumerate}[label=\arabic*) ]
      \item Switch to permeate mode, buffer drains to near-empty, then fills with \(\approx568\)~L permeate via \PUR.
      \item Perform the manual inversion on the evaporator.
      \item Switch back to concentrate mode, resume \GRN~band control from the same source silo.
    \end{enumerate}

    \textbf{Traceability:}
    All buffer fills (\PUR/\GRN), mode/transition events, per-silo desugaring volumes, inlet \brix\ readings, and line backflush actions are recorded with a unique \texttt{evap\_session\_id}. This allows reconstruction of mass balances and the exact boil order of source silos.

  \subsection{Permeate and utilities management}
    \label{subsec:permeate}

    The plant maintains a base permeate reserve in Silo~6, which is directly plumbed to the RO: it receives permeate during concentration and supplies permeate for RO cleaning. Silo~6 also feeds the \PUR~network pump for plant rinses and has a top overflow to drain (used only as a last resort). 
    At season start, a small reserve is created (e.g., RO on well water), after which the objective is to never fully empty Silo~6. Since its volume is constrained, dynamic silos allow maintaining larger reserves while multiple operations consume and generate permeate simultaneously. 

    Two complementary routines are used to maintain Silo~6 within a healthy band while maximizing usable permeate volume across the yard and minimizing overflow loss:
    
    \textbf{Topstock (Silo~6 high $\rightarrow$ push out).}
    When Silo~6 exceeds a high threshold \(H\) (e.g., 95\%) and a valid permeate destination exists from Section~\ref{subsec:dynamicsilos}, the CPS transfers permeate from Silo~6 to the selected dynamic silo via
    \[
    \text{Silo 6} \ \rightarrow \ \PUR \ \rightarrow \ \YEL \ \rightarrow \ \text{destination silo}
    \]
    The transfer stops when either the destination reaches capacity or Silo~6 falls to a target level \(T\) (e.g., 85\%).
    Using \(H>T\) provides hysteresis and prevents oscillation.
    
    \textbf{Downstock (Silo~6 low $\rightarrow$ pull back).}
    When Silo~6 drops below a low threshold \(L\) (e.g., 75\%) and at least one dynamic silo is tagged as permeate, the CPS replenishes Silo~6 by drawing from the dynamic silo that currently holds the \emph{least absolute volume} of permeate (not the lowest \%)\footnote{Selecting the least-volume donor for downstock reduces the risk of \textit{fragmentation} of the permeate reserve, (i.e., having multiple silos tagged as permeate with low volume), which would hinder the capacity to receive concentrate.}. 
    The routine ends when the source empties or Silo~6 reaches \(T\) (e.g., 85\%).
    Plumbing uses \RED~in recirculation configuration by closing the drain (V32) and Silo~6 output (V37), and opening \RED~to the inlet of \PUR~pump (V33):
    \[
    \text{source silo} \ \rightarrow \ \RED \ \rightarrow \ \PUR \ \rightarrow \ \YEL \ \rightarrow \ \text{Silo 6}
    \]

\section{Outcomes and discussion}
  \label{sec:outcomes}

  This section presents the operational outcomes from the deployment of the CPS as part of the production season of 2025, spanning from 2025-03-13 to 2025-04-28. The boiling center is located in La Patrie, Québec, centralizing production for 7 sites in a radius of 40 km, totalling $\approx$ 50,000 taps.
  
  \subsection{Productivity, labor, and flow stability}
    While the boiling-center throughput is fundamentally constrained by the production equipments (RO, evaporator, and filter presses), the CPS reduced labor intensity and the likelihood of human error. Delivery tasks, that previously required operators to manually sequence valves and remain in the manifold room to monitor the process for up to 30 minutes, are now armed once from the HMI with a live ETA, enabling operators to return to higher-value work while the system handles safe routing and cleaning (Section~\ref{subsec:delivery}). Manual valve sequences per delivery (about four sets of \(5\!-\!6\) manipulations, contingent on whether rinses were actually done) were eliminated; cleaning now always runs by design. For evaporator operation, the feed management (Section~\ref{subsec:evap}) removed frequent walks (15-20~s one way) from the evaporator to the valve room, reducing both distraction from boiling tasks and the tendency to rush decisions under time pressure. 

    The HMI also reduced procedural misses: for the 4-hour concentrate-boil inversion, the interface highlighted overdue status in red, replacing the prior manual tracking that was prone to lapses.
    
    Although isolated events (e.g., a single valve actuation replacing a 2-3~min manual changeover) are faster, their contribution to daily productivity is negligible compared to equipment limits. The main contribution to productivity is a steadier workflow with fewer interruptions and lower cognitive load, which was observed informally across the season.
    
    \subsection{Sanitary practices and contamination control}
    Because all paths are machine-actuated and the CPS has full awareness of segment contents, systematic post-transfer CIP rinses run consistently (Section~\ref{subsec:delivery}, Section~\ref{subsec:ro}, Section~\ref{subsec:evap}). Prior to deployment, small pockets of residual sugar at room temperature could stagnate and occasionally seed contamination when reopened; mid-season \emph{ropy} defects occurred at least once every year despite otherwise good practices. In the first season with the CPS, no contamination incident was reported.
    Late-season defects associated with warm inbound sap remain possible, but the baseline sanitary risk during slowdowns (e.g., deep freezes) appears effectively mitigated by the CIP implementation.
    
  \subsection{Traceability, auditability and data transparency}
    All deliveries were fully cataloged this season (\(n=150\)), including lot origin and full inline measurement data; delivery bills are generated automatically and emailed on completion. This immediacy eliminated the delayed feedback inherent to the previous manual process, which had led to information loss and client disputes about volume and/or sugar content. 
    Seven \emph{Exception} deliveries were isolated successfully.

    Administrative effort for reporting and payables was almost entirely eliminated ($\approx 1\text{h}$ for the whole season, compared to $> 30\text{h}$ across staff members before deployment). 
    This reduction is inherent to the automatic generation of delivery bills and cumulative reports, but also thanks to a drop in client disputes (informally reported as an approximate factor of 10) due to the instant, consistent documentation. 
    
  \subsection{Operability and safety}
    The centralized interlock queued \textbf{431} operations across the season without incident. While this number is inflated compared to the reality of the previous manual workflow, it shows that the CPS allowed to: (i) implement finer-grained process control, (ii) decrease operational risk from human error, (iii) absorb the increased complexity into system design to avoid incurring technical debt. 
    Hardware reliability was high: among 50 valve actuators, a single failure occurred (V01 -- evaporator concentrate feed). The issue was escalated and resolved on-site within \(\approx 15\) minutes by swapping in a spare actuator. During the swap, the buffer was maintained with permeate, causing only a minor slowdown (\(\approx 30\) minutes) instead of a potential multi-hour shutdown from evaporator starvation. 
    
    Permeate balancing (Section~\ref{subsec:permeate}) ran automatically throughout the season with \textbf{908} topstocks and \textbf{296} downstocks, maintaining reserves without operator intervention and minimizing overflow. Compared to the previous configuration employing fixed silos 5 and 6 (capacity 22,712 L) the maximum permeate total reserve observed after deployment was $\approx 41,640$ L, with 4/5 dynamic silos containing permeate.
    In the opposite direction, the maximum concentrate total reserve observed after deployment was $\approx 32,176$ L, with all dynamic silos containing concentrate. This shows the increase in flexibility gained from the dynamic silos, allowing both (i) more overall asset utilization and (ii) more total capacity for a given content type depending on the situation.
    
    The desugaring threshold (\(0.1~^\circ\)Bx) for \YEL~and \GRN~operations worked reliably; it should be noted that refractometer-based sensors can drift upward if not cleaned. The required cleaning interval was increased by the automated CIP operations (from $\approx$ 1 day to $\approx$ 2 weeks), yet manual cleaning was still necessary. For this reason, each desugaring operation enforces minimum/maximum runtimes as safeguards against improper execution caused by sensor drift.

    Network robustness was adequate: no plant-side control degradation occurred under temporary ISP loss (\(\approx 48\)~h) because orchestration and HMI run on a local server; only remote monitoring was affected. 

\section{Conclusion and future work}
\label{sec:conclusion}

  This paper presented a process-aware CPS piloting plant-level logistics, traceability and CIP in a maple syrup boiling center. Building on the architecture detailed in~\cite{bernardSugarShack402025}, four subsystems (delivery, RO, evaporator feed, permeate management) were implemented as state machines. These subsystems share an interlock-protected manifold and dynamic silos with priority-based destinations to maximize asset utilization. The result is a plant that executes safe, deterministic routing with consistent sanitary practices and complete, immediate traceability up to the evaporator inlet. While core throughput remains bounded by production equipment, the CPS demonstrably reduced labor intensity and error likelihood, eliminated mid-season contamination incidents, and enabled finer-grained operations (431 queued routines; 908 topstocks and 296 downstocks).

  Future work sees three complementary directions:
  \begin{itemize}
    \item \textbf{Extended process coverage}: integrate syrup output, filtering and barreling. By reaching full process coverage, the CPS would be able to produce a complete traceability chain for every barrel produced.
    \item \textbf{Reusable control assets}: package the routines/interlock patterns as Node-RED nodes/subflows (destination selection, desugaring, topstock/downstock, transition guards) with configuration schemas for rapid reuse; add site-to-site replication to evaluate generalizability across seasons and facilities.
    \item \textbf{Advanced analytics and robustness}: add online anomaly detection for drift in sensor readings, valve/pump actuation latencies; experiment integration of pH and viscosity data into syrup quality outcomes.
  \end{itemize}

  Beyond the boiling center, the same event-driven patterns (shared-resource arbitration, content-aware routing, and post-transfer hygiene) may also apply to other slow/medium-dynamic liquid agrifood processes (e.g., juice, brewing, dairy), where business constraints and traceability needs are similar.

\section*{Acknowledgments}
The authors would like to thank Justine Landry for her help with the hardware deployment and the development of the graphical interfaces, during her role as an intern.

\bibliographystyle{IEEEtran}
\bibliography{refs}

\begin{IEEEbiography}[{\includegraphics[width=1in,height=1.25in,clip,keepaspectratio]{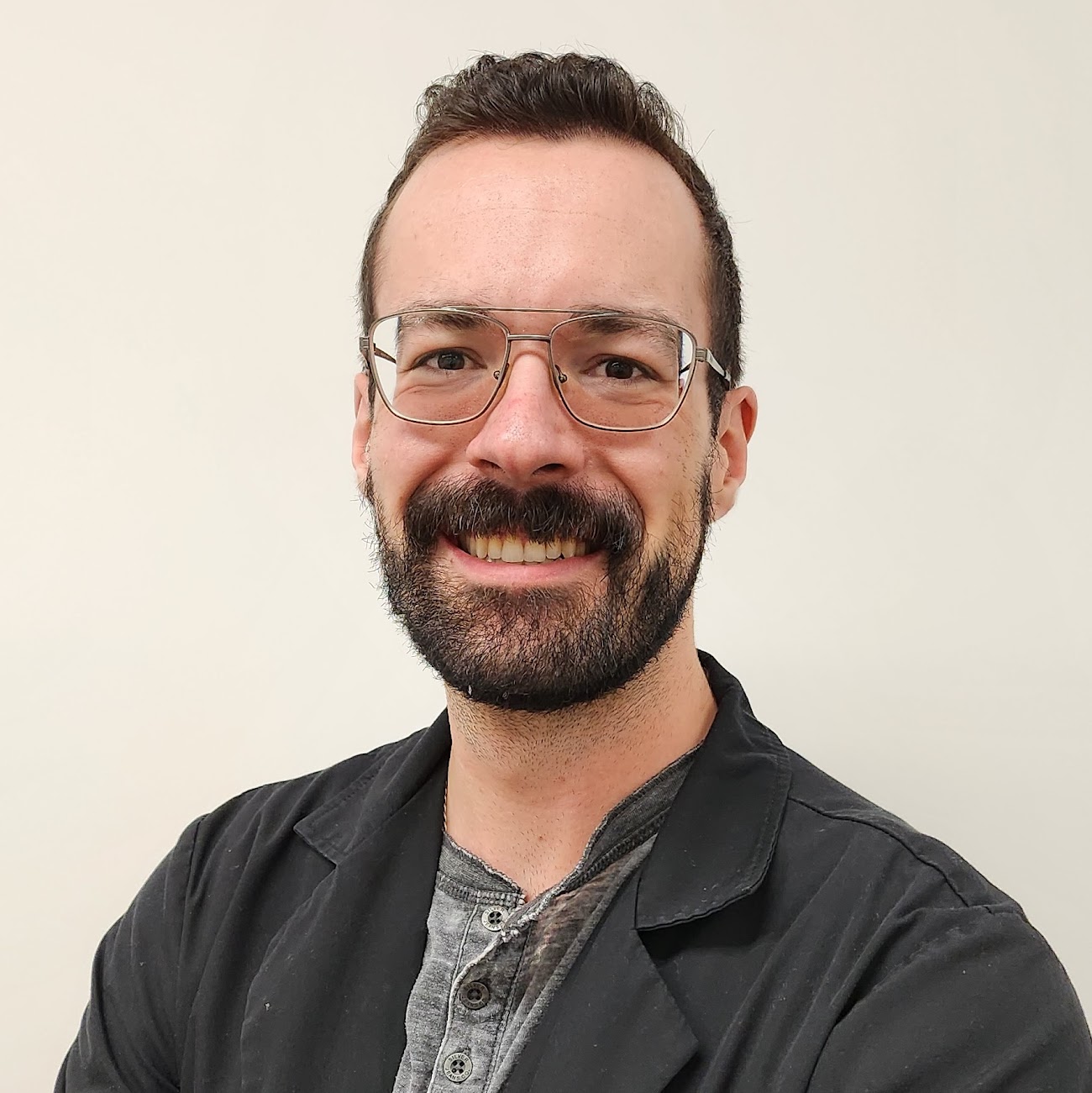}}]{Thomas Bernard} (Graduate Student Member, IEEE) received the B.Eng. degree in computer engineering in 2018, the M.Sc. degree in electrical engineering in 2021, and is currently pursuing the Ph.D. degree in electrical engineering, at Université de Sherbrooke, Sherbrooke, QC, Canada. His major field of study is Industry~4.0 automation systems.
  He has worked as an R\&D Consultant in industrial IoT since 2015, with roles spanning embedded systems, distributed automation, and data integration. He is currently R\&D Coordinator at AFCA, La Patrie, QC, and a Lecturer in Robotics Engineering at Université de Sherbrooke. He is currently a Candidate to the Engineering Profession (CEP) with the Ordre des Ingénieurs du Québec.
  \end{IEEEbiography}

  \begin{IEEEbiography}[{\includegraphics[width=1in,height=1.25in,clip,keepaspectratio]{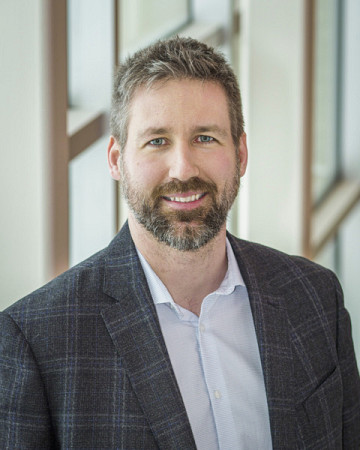}}]{François Grondin} (Member, IEEE) received the
      B.Sc. degree in electrical engineering from McGill
      University, Montreal, QC, Canada, in 2009, and the
      M.Sc. and Ph.D. degrees in electrical engineering
      from the Université de Sherbrooke, Sherbrooke,
      QC, Canada, in 2011 and 2017, respectively.
      After completing postdoctoral work with the
      Computer Science and Artificial Intelligence
      Laboratory, Massachusetts Institute of Technology,
      Cambridge, MA, USA, in 2019, he became a
      Faculty Member with the Department of Electrical
      Engineering and Computer Engineering, Université de Sherbrooke. He is currently a Member with the Ordre des ingénieurs du Québec.  His research interests include robot
      audition, sound source localization, speech enhancement, sound classification,
      and machine learning.
  \end{IEEEbiography}
  
  \begin{IEEEbiography}[{\includegraphics[width=1in,height=1.25in,clip,keepaspectratio]{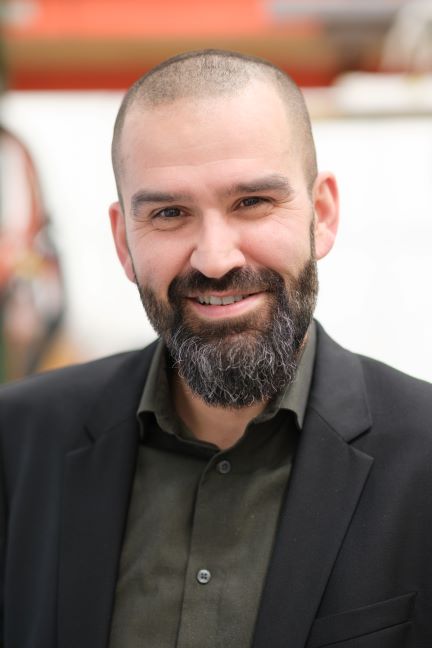}}]{Jean-Michel Lavoie}
    received the B.Sc. degree in chemistry and the M.Sc. and Ph.D. degrees in wood sciences from Laval University, Québec, Canada. Prof.~Lavoie is the founder and director of the Biomass Technology Laboratory, leading a team of 30--50 researchers dedicated to renewable energy and bioprocessing. His group collaborates with numerous industrial partners in Québec and abroad.
  \end{IEEEbiography}

\end{document}